\documentstyle[12pt,a4wide,epsf]{article}

\begin{document}

\begin{titlepage}
\renewcommand{\thefootnote}{\fnsymbol{footnote}}
\makebox[2cm]{}\\[-1in]
\begin{flushright}
\begin{tabular}{l}
DESY 94--172\\
TUM--T31--68/94\\
UAB--FT--348\\
hep--ph/9409440
\end{tabular}
\end{flushright}
\vskip0.4cm
\begin{center}
{\Large\bf Theoretical Update of the Semileptonic\\[4pt]
Branching Ratio of B Mesons}

\vspace{1.5cm}

E.\ Bagan$^1$,
Patricia Ball$^2$\footnote{Address after September 1994: CERN, Theory
Division, CH--1211 Gen\`{e}ve 23, Switzerland.},
V.M.\ Braun$^3$\footnote{On leave of absence from St.\ Petersburg
Nuclear Physics Institute, 188350 Gatchina, Russia.} and P. Gosdzinsky$^1$

\vspace{1.5cm}

$^1${\em Grup de F\'\i sica
Te\`orica, Dept.\ de F\'\i sica and Institut de F\'\i sica d'Altes
Energies, IFAE, Universitat Aut\`onoma de Barcelona,
E--08193 Bellaterra (Barcelona), Spain}\\[0.5cm]
$^2${\em Physik-Department/T30, TU M\"{u}nchen, D--85747 Garching,
Germany}\\[0.5cm]
$^3${\em DESY, Notkestra\ss\/e 85, D--22603 Hamburg, Germany}

\vspace{1cm}

{\em September 29, 1994}

\vspace{1cm}

{\bf Abstract:\\[5pt]}
\parbox[t]{\textwidth}{
We reconsider the prediction of the semileptonic branching ratio of
B mesons, using the recent calculation \cite{tum67} of the
radiative corrections with account for finite quark masses in
nonleptonic decays and taking into account $1/m_b^2$ corrections.
For the semileptonic branching ratio we obtain $B_{SL} =
(11.8\pm 1.6)\%$
using pole quark masses and $B_{SL} = (11.0\pm 1.9)\%$ using
running $\overline{\rm MS}$ quark masses. The uncertainty is dominated
by unknown higher order perturbative corrections.
We conclude that the present
accuracy of the theoretical analysis
 does not allow to state a
significant disagreement with the experimental results.
However, our  re-analysis of the decay $b\to ccs$
yields an increase of $(35\pm 11)\%$ due to
next-to-leading order corrections including mass dependent terms, which
further emphasizes the problem of the average charm quark content of the
final states in B decays.
}

\vspace{1.7cm}
{\em Submitted to Phys.\ Lett.\ B}
\end{center}
\end{titlepage}
\setcounter{footnote}{0}
\newpage

{\large\bf 1.} The theoretical description of inclusive weak decays of heavy
hadrons has made considerable progress over the recent years, see \cite{S}
for a review.
It could be shown that in the limit of infinite heavy quarks the decay
rate coincides with that of the corresponding free quark decay;
corrections to that limit are of nonperturbative origin and suppressed
by at least two powers in the heavy quark mass \cite{PLB}.
Today there is increasing confidence that QCD
predictions of heavy particle decays rest on a firm theoretical foundation.
In view of these apparent advances and
with the availability of new and more precise data on the
semileptonic branching ratio $B_{SL}$ of the B meson \cite{ARGUS},
the long felt discrepancy between its measured and its predicted
value becomes more and more baffling \cite{bigi,FWD94}. Over the last
years, the main efforts were concentrated on the determination of the
size of nonperturbative
power-suppressed corrections to the free quark decay, which, however,
 turned out to be small, of natural size
$\sim 1\,{\rm GeV}^2/m_b^2\sim 5\%$, and
cannot explain the experimental value of $B_{SL}$. Thus, it seems
timely to place more emphasis on {\em perturbative} radiative corrections to
the free quark decay, which since the well-known analysis by Altarelli
and Petrarca \cite{AP91} have not been receiving proper attention
in the literature.
In this letter we update the theoretical prediction of the semileptonic
branching ratio of B mesons using a recent calculation \cite{tum67}
of the charm quark mass dependence of radiative corrections to nonleptonic
decays. In addition we re-analyze the decay rate $\Gamma(b\to ccs)$,
taking into account the  quark mass dependence of radiative
corrections and the contributions of
penguins. The semileptonic branching ratio is then evaluated using
both pole masses and running
quark masses. The latter procedure  was advocated in \cite{BN}
on the evidence of the cancellation of renormalon singularities
 \cite{BBZ}. Finally, we discuss shortly the problem of fixing a proper
renormalization scale in heavy quark decays.

\bigskip

{\large\bf 2}. The semileptonic branching ratio of B mesons is defined by
\begin{equation}\label{eq:BRSL}
B_{SL} \equiv  \frac{\Gamma(B\to X e\nu)}{\sum_{\ell =
e,\,\mu,\,\tau}\!\Gamma(B\to X\ell\nu_{\ell}) + \Gamma(B\to
X_c) + \Gamma(B\to X_{c\bar c})+\Gamma({\rm rare\ decays})}.
\end{equation}
The heavy quark expansion (HQE) allows to relate the inclusive decay rate of
a B meson to that of the underlying b quark decay process, apart from
$1/m^2$ corrections:
\begin{equation}
\Gamma(B\to X) = \Gamma(b\to x) + {\cal O}(1/m_b^2).
\end{equation}
The power-suppressed correction terms to the total inclusive widths of
both semi-- and nonleptonic decays were
calculated in \cite{PLB,bigi}.

For the free quark decay rates we introduce the following notations:
\begin{eqnarray}
\Gamma(b\to c\ell\nu) & = & \Gamma_0\,{\rm
PH}(x_c,x_\ell,0)\,I(x_c,x_\ell,0),\\
\Gamma(b\to cud+cus) & = & 3\Gamma_0\,{\rm
PH}(x_c,0,0)\,\eta(\mu)\,J(x_c,\mu),\\
\Gamma(b\to ccs+ccd) & = & 3\Gamma_0\,{\rm
PH}(x_c,x_c,x_s)\,\kappa(x_c,x_s,\mu)\,K(x_c,x_s,\mu).\label{eq:kappaK}
\end{eqnarray}
Here $\Gamma_0$ is defined by $\Gamma_0 =
G_F^2|V_{cb}|^2m_b^5/(192\pi^3)$.
PH$(x_1,x_2,x_3)$ is the tree level phase  space factor of the decay
$b\to q_1+W\to q_1+\bar{q}_2+q_3$; for arbitrary masses
$x_i =m_i/m_b$ it is given by:
\begin{equation}
{\rm PH}(x_1,x_2,x_3) =
12\!\!\int\limits_{(x_2+x_3)^2}^{(1-x_1)^2} \! \frac{ds}{s}
\,(s-x_2^2-x_3^2)\,(1+x_1^2-s)\, w(s,x_2^2,x_3^2)\,w(s,x_1^2,1)
\end{equation}
with
\begin{equation}
w(a,b,c) = (a^2+b^2+c^2-2 ab-2ac-2bc)^{1/2}.
\end{equation}
The functions $\eta$ and $\kappa$ contain the leading-order
QCD corrections to the nonleptonic rates $b\to cuq$ and $b\to ccq$,
respectively. In particular, $\eta$ is given by \cite{gaillard}
\begin{equation}
\eta(\mu) =
\frac{1}{3}\left\{2\left(\frac{\alpha_s(m_W)}{\alpha_s(\mu)}
\right)^{4/\beta_0} +\left(\frac{\alpha_s(m_W)}{\alpha_s(\mu)}
\right)^{-8/\beta_0}\right\}
\end{equation}
with $\beta_0 = 11-2n_f/3$ for $n_f$ running flavours, $n_f=5$ in our
case.  The expression for $\kappa(\mu)$ is given below. Finally, $I$, $J$ and
$K$ contain the next-to-leading QCD corrections to the decay rates.
The function $I$ can be written as
\begin{equation}
I(x_1,x_2,x_3) = 1 +
\frac{2}{3}\,\frac{\alpha_s}{\pi}\,g(x_1,x_2,x_3),
\end{equation}
where $g$ has been calculated in Ref.\ \cite{HP} for
arbitrary arguments in terms
of an one-dimensional integral. Analytic expressions  are available for the
 special cases $g(x_1,0,0)$
 \cite{nir} and  $g(0,x_2,0)$ \cite{tum67}. The complete calculation of
$J(x_c,\mu)$ was first done in \cite{tum67}, while
$J(0,\mu)$ is also available from \cite{ACMP81}; some of the terms for
arbitrary $x_c$ have been also calculated in \cite{HP}.
 The function $K$ is not
yet known completely; we will discuss it below.

Summarizing existing calculations of the radiative corrections,
 we give the numerical values of
$g(x_c,0,0)$, $g(x_c,x_\tau,0)$ and $J(x_c,m_b)$ in Table~1. The
numbers are evaluated for $\alpha_s(m_Z)=0.117$,
i.e.\ $\Lambda_{\overline{\rm\scriptsize MS}}^{(4)}=312\,$MeV, and at
the renormalization scale $\mu=m_b=4.8\,$GeV. With these parameters,
we find $\eta(m_b)=1.10$.

\bigskip

{\large\bf 3}. As explained above, all the decay rates
entering the semileptonic branching ratio (\ref{eq:BRSL}) are known to
next-to-leading order in the strong interaction including final
state particle mass effects, except for $\Gamma(b\to ccs+ccd)$ and the
rare decays. Whereas the latter can safely be neglected, the channel
$b\to ccs$ deserves a closer consideration. In addition to the
contributions studied in Ref.\ \cite{tum67}, where
a 30\% increase of the decay rate $b\to ccs$ by radiative
corrections was obtained\footnote{See also \cite{vol1}.},
 we take into account the
 dependence of these
corrections on the s quark mass and discuss the contributions of
penguins.

The leading order decay rate can be written as
\begin{eqnarray}
\left.\vphantom{\sum}
\Gamma(b\to c\bar c s)\right|_{\rm\scriptsize LO} & = & 3\Gamma_0
|V_{cs}|^2 \, {\rm PH}(x_c,x_c,x_s)\left\{ \sum_{i=1}^6
c_i^2(\mu) + 2 \left[ \frac{1}{3}\,c_1(\mu) c_2(\mu) + c_1(\mu)
c_3(\mu)\right.\right.\nonumber\\
& & \left.+ \frac{1}{3}\, c_1(\mu)
c_4(\mu) + \frac{1}{3}\,c_2(\mu) c_3(\mu) + c_2(\mu) c_4(\mu) +
\frac{1}{3}\, c_3(\mu) c_4(\mu) + \frac{1}{3}\, c_5(\mu)
c_6(\mu)\right]\nonumber\\
& & {} - 2 f(x_c,x_c,x_s)\left[ c_1(\mu) c_5(\mu) +
\frac{1}{3}\, c_1(\mu) c_6(\mu) + \frac{1}{3}\, c_2(\mu) c_5(\mu) +
c_2(\mu) c_6(\mu)\right.\nonumber\\
& & \left.\left. + c_3(\mu) c_5(\mu)+ \frac{1}{3}\, c_3(\mu) c_6(\mu) +
\frac{1}{3}\, c_4(\mu) c_5(\mu) + c_4(\mu) c_6(\mu)\right]\right\}
\label{eq:bccs}\\
& \equiv & 3\Gamma_0|V_{cs}|^2 \, {\rm PH}(x_c,x_c,x_s)\,\kappa(x_c,x_s,\mu).
\label{eq:defkappa}
\end{eqnarray}
The coefficients $c_i(\mu)$, $1\leq i\leq 6$, are the leading order
Wilson-coefficients multiplying the operators $Q_i$ in
the effective Lagrangian and can be found in tabulated form in
\cite{buras}. Unlike the expression in Ref.\ \cite{AP91}, our
Eq.\ (\ref{eq:bccs})
also takes into account the interference of four-quark operators having
the usual $(V-A)\otimes (V-A)$ structure with penguin operators of the
structure $(V-A)\otimes (V+A)$. These interference terms are
explicitly of order $x_c^2$ and enter the decay rate with a
weight-function $f$, given by
\begin{equation}
f(x_c,x_c,x_s) = \frac{1}{{\rm
PH}(x_c,x_c,x_s)}\,\int\limits_{(x_c+x_s)^2}^{(1-
x_c)^2}\!\!ds\,\frac{6x_c^2}{s^2}\,w(s,x_c^2,x_s^2)\,w(1,s,x_c^2)\,
(s+x_s^2-x_c^2)\,(1+s-x_c^2).
\end{equation}
For reasonable quark
masses $x_c=0.3$ and $x_s=0.04$ we find $f=0.24$ and
$\kappa(\mu=m_b=4.8\,{\rm GeV})= 1.07$. Neglecting the
penguin-contributions, i.e.\ for
$c_i(\mu)\equiv 0$ for $i\geq 3$, $\kappa(m_b)$ coincides with
$\eta(m_b)=1.10$, so
that the penguins interfere destructively and
reduce the decay rate by $\sim 3\%$ similarly to what was
observed in \cite{AP91}.

In next-to-leading order, the decay rate can be written as in Eq.\
(\ref{eq:kappaK}), where $K$ is defined by
\begin{equation}
\kappa(x_c,x_s,\mu)K(x_c,x_s,\mu) \equiv \sum_{i,j=1}^6
f_{ij}(x_c,x_s) c_i(\mu)c_j(\mu) d_{ij}(x_c,x_s,\mu),
\end{equation}
the weight-factors $f_{ij}$ being given in (\ref{eq:bccs}), whereas
the $d_{ij}$ have the structure
\begin{equation}
d_{ij}=1+k_{ij}\,\frac{\alpha_s(\mu)}{\pi} +
r_{ij}\,\frac{\alpha_s(m_W)-\alpha_s(\mu)}{\pi} + {\cal O}(\alpha_s^2).
\end{equation}
The terms $r_{ij}$ contain matching-coefficients and two-loop
anomalous dimensions of the operators $Q_i$ and can be obtained from
\cite{buras}. The terms $k_{11}$ and $k_{22}$ can be obtained from
\cite{tum67,HP}, including all dependence on $x_c$ and $x_s$, likewise
$k_{12}$ for $x_c=x_s=0$ and, partly, also in
dependence on $x_c$ and $x_s$. The other terms
are not known. Nevertheless, the knowledge of these three
coefficients allows a rather accurate determination of the decay rate
to next-to-leading accuracy: for the unknown $d_{ij}$,
we most conservatively assume $0<d_{ij}<2$, which corresponds to
$|k_{ij}|<15$ for
$\alpha_s=0.2$. For $d_{12}$, we replace the uncalculated term,
$H_e(x_c,x_s)$ in the notation of \cite{tum67}, by its corresponding
value for only one massive c quark, $G_e(x_c)$. We estimate the error
introduced by this procedure by $\Delta H_e\approx 2
|G_e(x_c)-G_e(0)|$. The values of the relevant $k_{ij}$ are given in Table~2,
together with the functions $\kappa$ and $K$, the latter one yielding
the increase of $\Gamma(b\to ccs)$ due to next-to-leading order QCD
corrections.

At this point, it is worthwhile to emphasize that the dependence of the
decay rate on the s quark mass is rather weak. While the phase space
factor is considerably reduced by including the strange mass \cite{AP91},
this effect turns out to be to a large extent compensated by the increase
of radiative corrections. For $x_c=0.3$ we find that a
strange quark mass  $m_s=200$ MeV, $x_s=0.04$,
reduces the decay rate by 1.5\% only, which is smaller than the effect
of the penguins.

Taking everything together, we find that for
$x_c=0.3$ and $x_s=0.04$, next-to-leading order
radiative corrections increase $\Gamma(b\to ccs)$ by $(35\pm 7^{+8}_{-7})\%$,
where the first error is a very conservative estimate of the unknown parts
of the next-to-leading
corrections and the second error comes from a variation of
the renormalization scale $\mu$ within $m_b/2< \mu< 2 m_b$.

\bigskip

{\large\bf 4}. We have now all ingredients at hand to evaluate
$B_{SL}$. Before
doing so, however, let us make some remarks about the
nonperturbative corrections entering the decay rates. They can be
expressed in terms of two hadronic matrix elements, $\lambda_1$ and
$\lambda_2$.
Whereas $\lambda_2$ is directly related to the observable
spectrum of beautiful mesons,
\begin{equation}
\lambda_2 \approx \frac{1}{4}\,(m_{B^*}^2 - m_B^2)= 0.12\,{\rm GeV}^2,
\end{equation}
the quantity $-\lambda_1/(2m_b)$, which can be interpreted as the average
kinetic energy of the b quark inside the B meson,
is only difficult to measure.
At present, only a QCD sum rule calculation is available,
according to which $\lambda_1 = -(0.5\pm 0.1)\,$GeV$^2$ \cite{BB94}.
For a summary of the discussion about $\lambda_1$ we refer to
\cite{NPro}. The formulas for the decay rates including
power-suppressed corrections are given in \cite{bigi}.

We next have to fix the quark masses that enter the decay rates. For
the strange quark mass we use $m_s=0.2\,$GeV; as emphasized in the
last section, the decay rates are not very sensitive to this
parameter. As for the heavy quark masses, we make use of
the fact that in the framework of HQE the difference between
$m_b$ and $m_c$ is fixed:
\begin{equation}
m_b - m_c = m_B-m_D + \frac{\lambda_1+3\lambda_2}{2}\!\left(
\frac{1}{m_b} -\frac{1}{m_c} \right)
+ {\cal O}\!\left(\frac{1}{m^2_Q}\right)\! .\label{eq:gaehn}
\end{equation}
For $m_b$ we use $m_b=(4.8\pm0.2)\,$GeV. Varying the renormalization
scale $\mu$ within the range $m_b/2<\mu<2m_b$, we find
\begin{equation}\label{eq:hoho!}
B_{SL} = (11.8\pm 0.8\pm 0.5\pm 0.2\pm 0.2^{+0.9}_{-1.3})\%\,,
\end{equation}
which is our main result.
Here the first error comes from the uncertainty in $m_b$, the second
one from the one in $\alpha_s(m_Z)=0.117\pm 0.007$, the third one from
the uncertainty in $\lambda_1$ and the fourth from the uncertainty in
$\Gamma(b\to ccs)$. The last error comes from the variation of the
renormalization scale\footnote{ Larger values of $m_b$ and (or) of the
normalization scale generally yield a larger $B_{SL}$, while the increase
of $\alpha_s$ tends to lower the branching ratio.}.

The error stemming from the uncertainty in $\mu$ is rather big and
shows that higher order perturbative corrections are important. We
thus feel motivated to evaluate $B_{SL}$ also in a different scheme,
using running short--distance masses, e.g.\ $\overline{\rm MS}$
masses. This procedure has also been
advocated in connection with the cancellation of renormalon
contributions \cite{BBZ}. In
order to keep the formulas scheme-independent at ${\cal
O}(\alpha_s)$, the phase-space has to be modified according to
\begin{equation}
m_b^5\,{\rm PH}(x_c,0,0) \longrightarrow \bar{m}_b^5\,
{\rm PH}(\bar{x}_c,0,0)\left\{1+
\frac{\bar{\alpha}_s}{\pi}\left(\frac{20}{3}-5\ln\,\frac{\bar{m}_b^2}{\mu^2}-2
\bar{x}_c \ln \bar{x}_c\, \frac{d\ln{\rm
PH}(\bar{x}_c,0,0)}{d\bar{x}_c} \right)\right\},
\end{equation}
where $\bar{x}_i$ denotes running quantities evaluated at the scale
$\mu$. With this substitution, we obtain
\begin{equation}
\bar{B}_{SL} = (11.0\pm 0.6\pm 0.8\pm 0.2\pm 0.2^{+1.0}_{-2.2})\%
\end{equation}

In Table~3 we give a comparison of theoretical predictions for $B_{SL}$
using different approximations. The main result of our analysis is
that the prediction of Altarelli and Petrarca is lowered by more than 1\%.
It is clearly visible that the main effect comes from taking into account
the quark mass dependence of radiative corrections calculated in
\cite{tum67}, while the nonperturbative $1/m_b^2$ corrections
result in a $\sim 0.2\%$ decrease, in agreement with \cite{bigi}.

The results shown above clearly demonstrate that the main theoretical
uncertainty in the predictions for B decays comes from the scale and scheme
dependence of the results. As always, the only consistent way to reduce
 both is to make a next-to-next-to-leading (NNLO) calculation of the decay
rate including ${\cal O}(\alpha_s^2)$ corrections, which is a formidable
enterprise. Lacking this calculation, one is bound to make rather crude
estimates of the possible higher-order corrections using some particular
prescription to fix the scale. Among these, the BLM prescription \cite{BLM}
seems to us to be the only one that is physically motivated. We now
discuss in short its possible outcome on the B decay widths.

The idea underlying the BLM approach is that the major part of
higher-order radiative corrections originates from the necessity to
evaluate Feynman diagrams with the running coupling at the scale of
the gluon virtuality and can be traced by a relatively simple calculation
of the diagrams with an extra fermion bubble in the gluon line.
The corresponding calculations have been done recently \cite{VS,luke} and
indicate that the natural scale in the radiative corrections in
B decays is significantly smaller than $m_b$. In particular,
neglecting the c quark mass, Ref.\ \cite{luke}
finds $\mu_{BLM}=0.07 m_b$
for the radiative corrections to the semileptonic
decay,\footnote{ Note that this result is
contrary to expectations in \cite{bigi}, where the choice of a low
scale in the radiative corrections was criticized.}
while for the final state interaction of quarks in the nonleptonic
decays  $\mu_{BLM}\sim 0.32 m_b$ is found, as indicated
by the studies of the $\tau$ lepton hadronic decay width.
In fact, the particular scale entering the
radiative corrections to the semileptonic width turns out to be not very
important for the problem of the semileptonic branching ratio, since these
corrections cancel to a large extent between numerator and denominator
of Eq.\ (\ref{eq:BRSL}).\footnote{In the case at hand, where the BLM
prescription
indicates a {\em very} low scale, we find it more appropriate not to change
the scale $\mu=m_b$ to $\mu = 0.07m_b$, but rather to use the
calculation in \cite{luke}
as an explicit estimate of the $\alpha_s^2(m_b)$ correction.}
The main problem in applying the BLM scheme to nonleptonic B decays is
that the radiative corrections in NNLO do not
factorize into ``semileptonic'' and ``final-state-interaction''
parts. On may thus suspect that a large part of the radiative corrections
comes from other types of diagrams than those considered
in the BLM method (see the discussion in \cite{vol1}). In addition, it
is not clear how to apply this approach consistently to processes where
the relevant operators possess a nontrivial anomalous dimension.
Still, we believe that the low scales indicated by the BLM prescription are
more natural in B decays. Thus the choice $\mu=2 m_b$ adopted above as one
extreme case is in fact rather
unlikely, while $\mu=m_b/2$ is presumably more relevant. Adopting this scale,
our result for the semileptonic branching ratio in (\ref{eq:hoho!}) becomes
$B_{SL}=(10.5\pm 1.4)\%$, in perfect agreement with the experimental
value $B_{SL}^{\rm\scriptsize exp} = (10.4\pm 0.4)\%$ \cite{ARGUS}.
Summarising, we conclude that there is no evidence for any disagreement
between the experimental data and the theoretical prediction for the
semileptonic branching ratio of B mesons.

The situation is not so clear, however, with the charm content in the
final states. With the 35\% increase of the $b\to c\bar c s $ rate induced
by taking into account the c quark mass in the radiative corrections,
this problem is strengthened. From our analysis we get
\begin{equation}\label{eq:aetsch}
\langle\,n_c\,\rangle = 1.28\pm 0.08,
\end{equation}
which is to be confronted with the experimental result $\langle\,n_c\,
\rangle^{\rm\scriptsize exp} = 1.04\pm 0.07$ \cite{roudeau}.
We are not aware of any natural theoretical possibility
to lower the value given in (\ref{eq:aetsch}),
unless the c quark mass is much larger than expected, which would
conflict, however, with the heavy quark expansion of the meson
masses, Eq.\ (\ref{eq:gaehn}).

\newpage

\renewcommand{\textfraction}{0}
\renewcommand{\arraystretch}{1.2}
\addtolength{\arraycolsep}{5pt}
\section*{Tables}
\begin{table}[h]
$$
\begin{array}{l|ccc}
x_c & g(x_c,0,0) & g(x_c,x_\tau,0) & J(x_c,m_b)\\
\hline
0   & -3.62 & -3.37 & 1.009 \\
0.1 & -3.25 & -2.89 & 1.026 \\
0.2 & -2.84 & -2.42 & 1.046 \\
0.3 & -2.51 & -2.08 & 1.063 \\
0.4 & -2.23 & -1.81 & 1.077 \\
0.5 & -2.01 & -1.61 & 1.088 \\
0.6 & -1.83 & -1.45 & 1.097 \\
0.7 & -1.70 &       & 1.105 \\
0.8 & -1.59 &       & 1.113 \\
0.9 & -1.53 &       & 1.123 \\
1   & -1.50 &
\end{array}
$$
\caption[]{Next-to-leading order corrections to the semileptonic b
quark decay rates and the decay $b\to cud$ as functions of
$x_c=m_c/m_b$. Parameters: $\mu=m_b=4.8\,$GeV,
$\Lambda_{\overline{\rm\scriptsize MS}}^{(4)} = 312\,$MeV,
corresponding to $\alpha_s(m_Z) = 0.117$; $x_\tau=m_\tau/m_b$.
Note that $J(1,\mu)$ diverges like $\sim\ln (1-x_c)$.}\label{tab:1}
\end{table}
\begin{table}[h]
$$
\begin{array}{l|ccccc}
x_c & \kappa(x_c,x_s,m_b) &k_{11} & k_{12}(\mu=m_b) & k_{22} &
K(x_c,x_s,m_b)\\ \hline
0 & 1.054 & -1.33 & -7.59\pm 0.01 & -1.26 & 1.02\pm 0.05\\
0.1 & 1.056 & -0.05 & -6.65\pm 0.07 & -0.35 & 1.09\pm 0.06\\
0.2 & 1.062 & \phantom{-}2.53 & -4.97\pm 0.20 & \phantom{-}1.23
& 1.20\pm 0.06\\
0.3 & 1.069 & \phantom{-}6.69 & -2.64\pm 0.57 & \phantom{-}3.41
& 1.35 \pm 0.07\\
0.4 & 1.077 & 15.68 & \phantom{-}1.24\pm 0.96 & \phantom{-}7.09
& 1.62\pm 0.09
\end{array}
$$
\caption[]{The leading and next-to-leading order corrections to
the nonleptonic
decay $b\to ccs$. The errors rely on a conservative estimate of
the unknown parts of the next-to-leading order terms, mostly due to penguin
contributions. The last column gives the increase of the decay rate
$\Gamma(b\to ccs)$ in next-to-leading order including finite c and s
quark effects in the
radiative corrections. The input parameters are the same as in
Table~1; $x_s=0.04$.}\label{tab:2}
\end{table}
\begin{table}[h]
\begin{tabular}{c|cccc}
& Parton Model \cite{AP91} & HQE \cite{bigi} & \multicolumn{2}{c}{HQE
[this work]}\\
$\alpha_s(m_Z)$ & pole masses & pole masses & pole masses &
$\overline{{\rm MS}}$ masses\\ \hline
0.110 & 0.133 & 0.132 & 0.123 & 0.117\\
0.117 & 0.130 & 0.128 & 0.118 & 0.110\\
0.124 & 0.125 & 0.123 & 0.113 & 0.102
\end{tabular}
\caption[]{
Theoretical predictions for the semileptonic branching ratio $B_{SL}$
depending on $\alpha_s(m_Z)$. Input parameters: $m_b=4.8\,$GeV,
$m_c = 1.33\,$GeV (pole masses), corresponding to $\lambda_1 = -0.5\,$GeV$^2$,
$m_s=0.2\,$GeV. Renormalization scale: $\mu = m_b$.}\label{tab:3}
\end{table}

\clearpage

\setcounter{table}{2}
\makebox[2cm]{}\\[-1in]
\begin{flushright}
\begin{tabular}{l}
February 1996
\end{tabular}
\end{flushright}
\vskip0.2cm
\begin{center}
{\large\bf Erratum:}\\
{\Large\bf Theoretical Update of the Semileptonic\\[4pt]
Branching Ratio of B Mesons\\[4pt]
[Phys.\ Lett.\ B {\bf 342} (1995) 362]}
 
\vspace{0.5cm}
 
E.\ Bagan, Patricia Ball, V.M.\ Braun and P. Gosdzinsky
 
\end{center}

\vspace{0.5cm}

In the third line in Eq.~(10) on page 364 there is a sign error:
$-2f(x_c,x_c,x_s)$ should read $+2f(x_c,x_c,x_s)$. In addition, we
have found an
error in the computer program, which affected the average charm
content $n_c$ and
the scale dependence of the results in the $\overline{\rm MS}$ scheme.
We take this opportunity to incorporate the complete results for the
quark mass
dependence of the radiative corrections to $b\to ccs$ calculated in
[22]. The corresponding update of our Table 2 on page 365 is 
given in Table 2 in [22].

The numerical impact of these corrections is marginal:
Eqs.~(17) and (19) on page 366 should read:
$$
B_{SL} = (12.0\pm 0.7 \pm 0.5 \pm 0.2^{+0.9}_{-1.2})\%\,,  \eqno (17)
$$
$$
\bar{B}_{SL} =  (11.3 \pm 0.6 \pm 0.7 \pm 0.2^{+0.9}_{-1.7})\%\,.\eqno (19)
$$
Table 3 on page  367 has to be replaced by the Table given below.

Since the problem of the average charm content is receiving increasing
attention (see, e.g.[23]), we give the corrected result for
$n_c$ in a somewhat expanded form. Eq.~(20) on page 367 is to be 
substituted by 
$$
n_c = 1.24 \pm 0.05 \pm 0.01\,,\eqno (20)
$$
which shows the result in the OS scheme.
The first error comes from the uncertainty in $m_b = (4.8\pm
0.2)\,$GeV, the second one from the uncertainties in the
remaining parameters.
In the $\overline{\rm MS}$ scheme we get
$$
\bar{n}_c = 1.30 \pm 0.03 \pm 0.03 \pm 0.01\,,\eqno (20')
$$
where again the first error comes from the uncertainty in the quark masses,
the second one is due to the variation of $\alpha_s$, and the third one
comprises the remaining uncertainties.

We have added a figure showing the charm content
versus the semileptonic branching ratio, cf.~[23],
obtained by relaxing the constraint 
on the quark masses following from the 
heavy quark expansion in Eq.~(16) on page
366 and allowing for a larger range of the ratio $m_c/m_b$.
Note that $m_c/m_b$ is scale-independent;  both $n_c$ and $B_{SL}$  
are functions of $m_c/m_b$, $\mu$ and $\alpha_s(\mu)$.

\vspace{0.5cm}

\noindent {\bf Acknowledgement:} We thank G.Buchalla and M. Neubert
for pointing out the errors.

\setlength{\parskip}{10pt}
\setlength{\parindent}{10pt}
\newpage
\noindent{\Large\bf References}

[22] E. Bagan {\em et al.}, Phys.\ Lett.\ B {\bf 351} (1995) 546.

[23] G. Buchalla, I. Dunietz and H. Yamamoto, Phys.\ Lett.\ B 
{\bf 364} (1995) 188.

[24] T.E.\ Browder, Talk given at {\em International
Europhysics Conference on High Energy Physics (HEP 95)}, Brussels
(Belgium), July 1995 (Hawaii Preprint PRINT-95-241);\\
T. Skwarnicki, Rapporteur Talk at {\em International
    Symposium on Lepton Photon Interactions (IHEP)}, Beijing
  (P.R.~China), August 1995 (hep-ph/9512395).
 
\vspace{0.5cm}

\begin{table}[h]
\begin{center}
\begin{tabular}{c|cccc}
& Parton Model [7] & HQE [5] & \multicolumn{2}{c}{HQE
[this work]}\\
$\alpha_s(m_Z)$ & pole masses & pole masses & pole masses &
$\overline{{\rm MS}}$ masses\\ \hline
0.110 & 0.133 & 0.132 & 0.124 & 0.120\\
0.117 & 0.130 & 0.128 & 0.120 & 0.113\\
0.124 & 0.125 & 0.123 & 0.114 & 0.105
\end{tabular}
\end{center}
\caption[]{
Theoretical predictions for the semileptonic branching ratio $B_{SL}$
as a function of $\alpha_s(m_Z)$. Input parameters: $m_b=4.8\,$GeV,
$m_c = 1.33\,$GeV (pole masses), corresponding to $\lambda_1 = -0.5\,$GeV$^2$,
$m_s=0.2\,$GeV. Renormalization scale: $\mu = m_b$.}
\end{table}
\begin{figure}[h]
\centerline{\epsffile{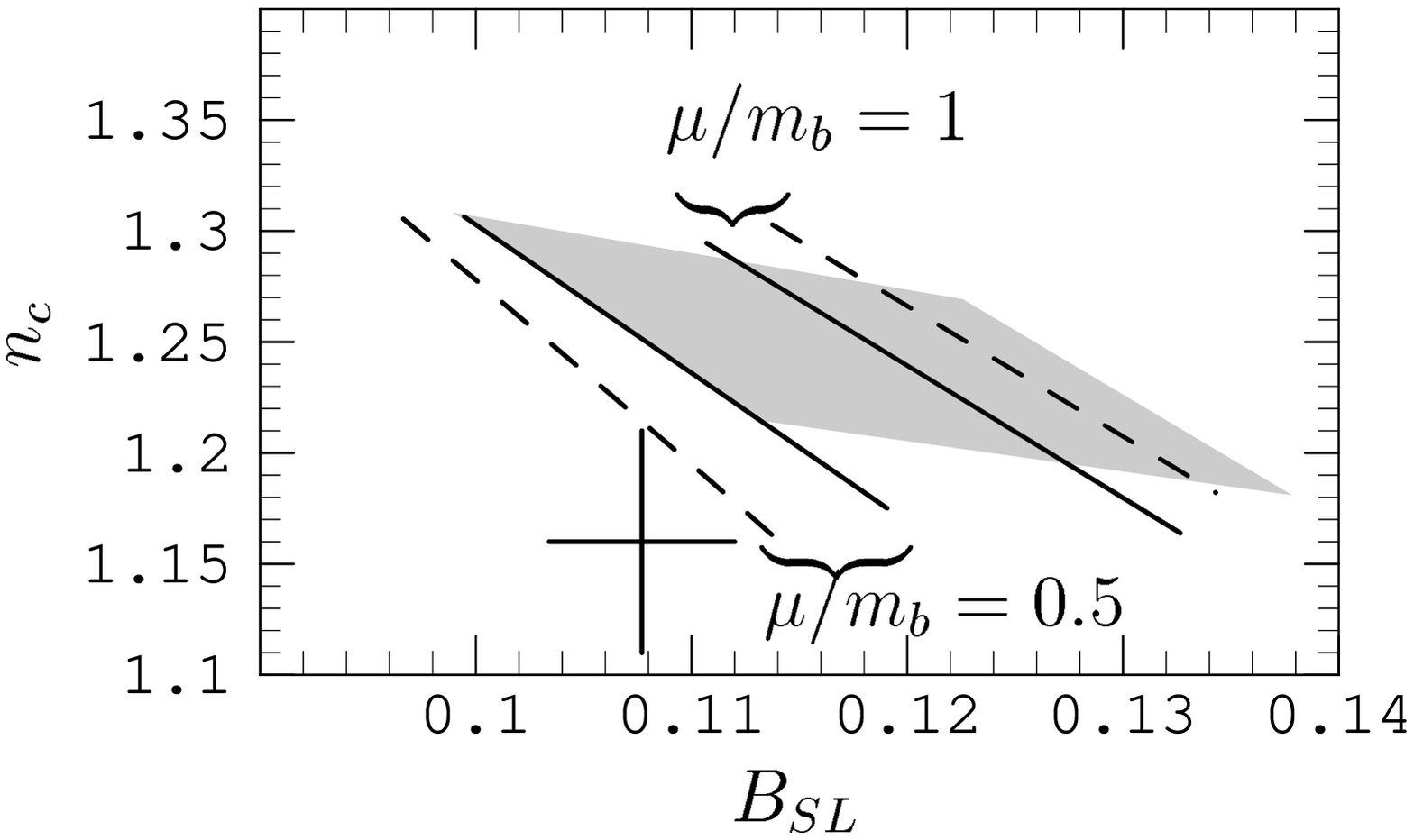}}
\caption[]{The charm content $n_c$ vs.\ $B_{SL}$. 
   Solid lines: theoretical predictions in the OS scheme for $0.23 <
  m_c/m_b < 0.33$, dashed lines: the same in the $\overline{\rm MS}$
  scheme for $0.18 < \bar{m}_c(\bar{m}_c)/\bar{m}_b(\bar{m}_b) <
  0.28$. 
  Shaded area: theoretical predictions in the OS scheme with
  $m_c$ obtained from Eq.~(16) and varying $m_b$, $\lambda_1$ and
  $\mu$ within the range of values given in the text.
  The experimental data point is taken from [24].}
\end{figure}

\newpage

\begin{thebibliography}{99}

\bibitem{tum67} E. Bagan {\em et al.}, TU M\"unchen Preprint
TUM--T31--67/94/R (hep--ph/9408306) (to appear in Nucl.\ Phys.\ {\bf B}).

\bibitem{S}
M.A. Shifman,
Talk given at Workshop on Continuous Advances in QCD, Minneapolis,
Preprint TPI--MINN--94--17--T (hep--ph/9405246).

\bibitem{PLB}
I. Bigi, N. Uraltsev and A. Vainshtein, Phys.\ Lett.\ B {\bf 293}
(1992) 430; Erratum {\em ibid.} {\bf 297} (1993) 477.

\bibitem{ARGUS}
R. Patterson, {\em Weak and Rare Decays}, Rapporteur Talk at ICHEP 94,
Glasgow, 20--27 July 1994;\\
M. Aguilar-Benitez {\em et al.} (Particle Data Group),
Phys.\ Rev.\ D {\bf 50} (1994) 1173.

\bibitem{bigi} I. Bigi {\em et al.}, Phys.\ Lett.\ B {\bf 323}
(1994) 408.

\bibitem{FWD94} A.F.\ Falk, M.B.\ Wise and I. Dunietz, CALTEC Preprint
CALT--68--1933 (1994) (hep--ph/9405346).

\bibitem{AP91} G. Altarelli and S. Petrarca, Phys.\ Lett.\ B {\bf 261}
(1991) 303.

\bibitem{BN} P. Ball and U. Nierste, TU M\"unchen Preprint
TUM--T31--56/94/R (hep--ph/9403407) (to appear in Phys.\ Rev. {\bf D}).

\bibitem{BBZ}
 I. Bigi {\em et al.}, Phys.\ Rev.\ D {\bf 50}
(1994) 2234;\\
M. Beneke and V.M.\ Braun, Nucl.\ Phys.\ {\bf B426} (1994) 301;\\
M. Beneke, V.M.\ Braun and V.I.\ Zakharov,
MPI M\"unchen Preprint MPI--PhT/94--18 (hep-ph/9405304).

\bibitem{gaillard} M.K.\ Gaillard and B.W.\ Lee, Phys.\ Rev.\ Lett.\
{\bf 33} (1974) 108;\\
G. Altarelli and L. Maiani, Phys.\ Lett.\ B {\bf 52} (1974) 351.

\bibitem{HP} Q.\ Hokim and X.Y.\ Pham, Phys.\ Lett.\ B {\bf 122}
(1983) 297; Ann.\ Phys.\ {\bf 155} (1984) 202.

\bibitem{nir} Y. Nir, Phys.\ Lett.\ B {\bf 221} (1989) 184.

\bibitem{ACMP81} G. Altarelli {\em et al.}, Nucl.\ Phys.\ {\bf B187}
(1981) 461;\\
G. Buchalla, Nucl.\ Phys.\ {\bf B391} (1993) 501.

\bibitem{vol1} M.B.\ Voloshin, Minneapolis Preprint
TPI--MINN--94/35--T (revised version) (hep--ph/9409391).

\bibitem{buras} A.J.\ Buras {\em et al.}, Nucl.\ Phys.\ {\bf B370}
(1992) 69.

\bibitem{BB94} P. Ball and V.M.\ Braun, Phys.\ Rev.\ D {\bf 49} (1994)
2472.

\bibitem{NPro} M. Neubert, Talk given at QCD 94, Montpellier, France,
7--13 July 1994.

\bibitem{BLM} S.J.\ Brodsky, G.P.\ Lepage and P.B.\ Mackenzie,
Phys.\ Rev.\ D {\bf 28} (1983) 228.

\bibitem{VS}
B.H.\ Smith and M.B.\ Voloshin,
Minneapolis Preprints UMN--TH--1241/94 (hep--ph/9401357);
UMN--TH--1252/94 (hep--ph/9405204).

\bibitem{luke} M. Luke, M.J.\ Savage and M.B.\ Wise, Toronto Preprint
UTPT--94--24 (hep--ph/9409287).

\bibitem{roudeau} P. Roudeau, {\em Heavy Quark Physisc}, Rapporteur
Talk at ICHEP 94, Glasgow, 20-27 July 1994.

\end{thebibliography}
\end{document}